\documentclass[fleqn,10pt]{wlscirep}
\usepackage{url}
\usepackage{color}
\usepackage{float}

\usepackage[utf8]{inputenc}
\usepackage[T1]{fontenc}
\usepackage{setspace}

\onehalfspace

\usepackage{setspace}

\title{Academic mentees succeed in big groups, \\but thrive in small groups}

\author[1,2]{Yanmeng Xing}
\author[1]{Ying Fan}
\author[2,3,4,5*]{Roberta Sinatra}
\author[1,$\dagger$]{An Zeng}
\affil[1]{School of Systems Science, Beijing Normal University, Beijing, P.R. China}
\affil[2]{Networks, Data, And Society (NERDS) Research Group, IT University of Copenhagen, 2300 Copenhagen, Denmark}
\affil[3]{Center for Social Data Science (SODAS), University of Copenhagen, 1353 Copenhagen, Denmark}
\affil[4]{ISI Foundation, 10126 Turin, Italy}
\affil[5]{Complexity Science Hub, 10126 Vienna, Austria}

\affil[*]{rsin@itu.dk}
\affil[$\dagger$]{anzeng@bnu.edu.cn}

\begin{abstract}
Mentoring is a key component of scientific achievements, contributing to overall measures of career success for mentees and mentors. A common success metric in the scientific enterprise is acquiring a large research group, which is believed to indicate excellent mentorship and high-quality research. However, large, competitive groups might also amplify dropout rates, which are high especially among early career researchers. Here, we collect longitudinal genealogical data on mentor-mentee relations and their publication, and study the effects of a mentor’s group on future academic survival and performance of their mentees. We find that mentees trained in large groups generally have better academic performance than mentees from small groups, if they continue working in academia after graduation. However, we also find two surprising results: Academic survival rate is significantly lower for (1) mentees from larger groups, and for (2) mentees with more productive mentors. These findings reveal that success of mentors has a negative effect on the academic survival rate of mentees, raising important questions about the definition of successful mentorship and providing actionable suggestions concerning career development.

\end{abstract}

\begin{document}
\maketitle
\flushbottom
\section*{Introduction}\label{sec:introduction}
Mentorship is fundamental in many professions\cite{scandura1992mentorship,de2003mentor,payne2005longitudinal,janosov2020elites}. In science, successful mentoring is crucial not only for a mentee's growth and success, but also for the career advancement of the mentor\cite{kram1988mentoring,lee2007nature,bhattacharjee2007nsf}. In mentoring relationships, mentees learn scientific values, skills, and build their scientific network\cite{green1991professional}. Also, mentorship has been shown to play a prominent role on a researcher's first job placement\cite{enders2005border,hilmer2007dissertation,wright2008mentoring,clauset2015systematic,way2019productivity}. At the same time, mentors obtain benefits from training mentees, like higher productivity, job satisfaction, and a broader social network in the long run\cite{allen2004career,astrove2017mentors}. A mentor can have multiple mentees over a career, and their number and success can improve the mentor's institutional recognition\cite{rossi2017genealogical,semenov2020network}. Yet, despite the important role of mentees and of junior researchers in the scientific ecosystem, we witness a large fraction of early-stage researchers exiting academia at an alarming rate\cite{roach2010taste,petersen2012persistence,moss2012science,ghaffarzadegan2015note,milojevic2018changing,xing2019strong,woolston2019phds,huang2020historical,levine2020covid,davis2022pandemic}, and we still have a limited quantitative understanding of the impact of mentors and their research group on the survival rate and career evolution. Given also the increasing reports of unhealthy working environments experienced by graduate students and early career researchers in academia\cite{levecque2017work, guthrie2018understanding, woolston2019phds, gonzalez2020risk, murguia2022navigating}, it is of fundamental importance to understand which kind of mentorship minimizes dropout rate, supports junior researchers' well-being, and enables talent diffusion.

The success of a mentor-mentee relationship is characterized by the complex interaction of different factors, like institutional environment, country of origin of the mentor and mentee, or funding for PhD research\cite{sugimoto2011academic,baruffaldi2016productivity,brostrom2019academic,way2019productivity}. Previous research on mentorship has been primarily based on anecdotal studies and self-report surveys, and supports the hypothesis that both mentees and mentors benefit from the mentoring relationship\cite{lewis1992carl,payne2005longitudinal}. Most recently, some large-scale quantitative studies provided a quantitative understanding of the interplay between mentor and mentee performance \cite{malmgren2010role, liu2018understanding, fortunato2018science}. For example, in Mathematics a mentor's fecundity, that is the number of mentees that a mentor supervises, is positively correlated with the number of the mentor's publications, and mathematicians have an academic fecundity similar to that of their mentors\cite{malmgren2010role}. Mentees in STEM fields not only learn technical skills and traditional knowledge\cite{liu2018understanding} but also inherit hidden capabilities displaying a higher propensity for producing prize-winning research, becoming a member of the National Academy of Science, and achieving ``superstardom''\cite{ma2020mentorship}. Researchers have a higher probability of continuing in academia if they can better synthesize intelligence between their graduate and postdoctoral mentors\cite{lienard2018intellectual,wuestman2020genealogical}. Moreover, graduate mentors are less instrumental to their mentees' survival and fecundity than postdoctoral mentors\cite{lienard2018intellectual}. However, mentees who show independence from the mentor's research topics after graduation have a higher tendency to be part of the academic elite\cite{ma2020mentorship}. Early-career investigators who coauthor with high impact scientists in the early stage have a long-lasting competitive advantage over those who do not have these collaboration opportunities\cite{li2019early}. Mentorship is also connected to the chaperone effect in scientific publishing\cite{sekara2018chaperone}: publishing with a senior mentor in a journal is crucial to become corresponding author in a later publication in the same journal, and this effect is particularly pronounced for high-impact publishing venues. 

These prior works have well demonstrated the positive association between the success of mentors and mentees. However, they have a major limitation: they mainly focus on the career success of those surviving in academia. As such, they are affected by survival bias and fail to capture why a mentee does not continue their academic career. Indeed, in a mentorship relation, a mentee can benefit from a mentor's broad vision and valuable research experience, especially when working with academically successful mentors. However, the mentee may face a strong competition for the mentor's limited time, since successful mentors tend to supervise more mentees, work with more collaborators\cite{johnson2002toward}, do more academic service, like peer reviewing or covering scientific editorial roles\cite{ma2020mentorship}, and manage scientific groups of large size \cite{luckhaupt2005mentorship,malmgren2010role,brostrom2019academic}. Therefore, mentees in large groups have to compete for the mentor's attention and have on average less chances of interactions with the mentor than mentees in small groups, entailing potential risks for the mentees' career evolution. 

Given this premise, here we ask a fundamental question: What are the advantages and disadvantages of working with successful mentors, especially in relation to their scientific group size? To address this question, we construct a dataset combining mentor-mentee relations and their academic records. This dataset can capture their academic proliferation and publication performance and can be used to explore the potential drivers of mentee success in academia\cite{ke2022dataset,david2012neurotree,sinha2015overview,wang2019review}. Most importantly, we can perform a survival analysis accounting for survivor bias, and understand not only the factors associated with success, but also those that lead to dropout. We further apply a coarsened exact matching regression model to uncover the causal relationship between mentees' group size and academic performance, which rules out potential confounding factors and uncovers alternative predictors\cite{iacus2009cem,iacus2012causal}. 

\section*{Results}\label{sec:result}
\subsection*{Data and data curation.}
Our analysis is based on two distinct data sets. The first one is curated from the Academic Family Tree (AFT, Supplementary S1.1), an online website  (\url{Academictree.org}) for collecting mentor-mentee relationship in a crowd-sourced fashion. AFT initially focused on Neuroscience and expanded later to span more than 50 disciplines. The second data set is the Microsoft Academic Graph (MAG, \url{https://aka.ms/msracad}, Supplementary S1.2), a bibliographic database containing entities about authors, doctypes (journals, conferences, etc.), affiliations, and citations. One advantage of MAG over other publication databases is that all entities have been disambiguated and associated with identifiers. These two data sets have been connected by matching the same scientists in each data set, and this matching has been validated with extensive and strict procedures\cite{ke2022dataset}. From this combined dataset, we extract the genealogical data of 309,654 scientists who published 9,248,726 papers in Chemistry, Neuroscience, and Physics between 1900 and 2021 (Methods and Supplementary Note 1 for data curation).

\begin{figure}[!bt]
    \centering
    \includegraphics[width=1\textwidth]{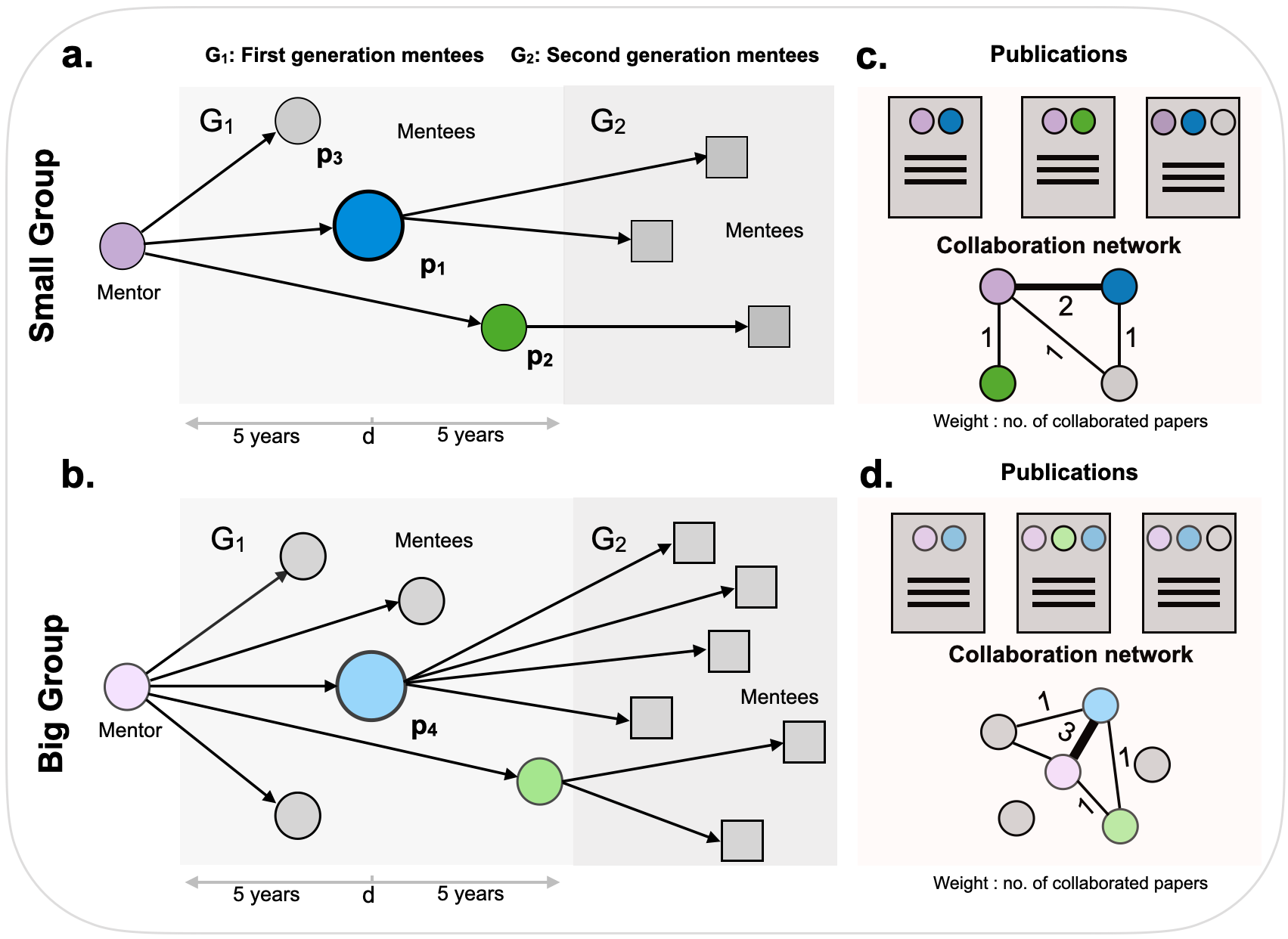}
    \caption{\textbf{Illustration of two academic family networks and mentor-mentee collaboration networks.} \textbf{a.} Genealogy network of a mentee. The network is built around a focal mentee $p_1$ (the large blue circle). The purple node corresponds to $p_1$'s mentor. A directed link between two nodes indicates the mentorship relation, where the mentor is the node the link departs from. $G_1$ indicates the set of nodes co-mentored by $p_1$’s mentor 5 years before and after $p_1$’s graduation. This time window is denoted as $d$. The squared nodes in $G_2$ are the mentees mentored by the nodes of $G_1$ during the first 25 years after their graduation. The blue node $p_1$ and the green node $p_2$ in $G_1$ have their mentees in $G_2$, whereas the grey node $p_3$ has no offspring, hence no mentee-node in $G_2$. Therefore, $p_1$ and $p_2$ are survived mentees, according to our definition, while $p_3$ drops out. Because of the number of nodes in $G_1$, $p_1$ is a mentee in a small group. \textbf{b.} Genealogy network of the mentee $p_4$. The difference in respect to panel a) is that the number of nodes in $G_1$, i.e. the group size, is among the top 25\% in the dataset, meaning that the mentee is mentored in a big group. \textbf{c.} Mentor-mentee co-authored publications and the corresponding weighted collaboration network among the mentor and mentees in the small group of panel a. Here the mentor and the $G_1$ mentees have co-authored three papers during the tranining period, resulting in a collaboration network where each node corresponds to an author and the edge weights represent the number of co-authored papers. \textbf{d.} Mentor-mentee co-authored publications and the corresponding weighted collaboration network among the mentor and their mentees in the big group of panel b.}
    \label{fig:schematic}
\end{figure}

\subsection*{Genealogy networks, mentee generations, and group size}
These curated datasets allow us to construct for each researcher $p$ their academic genealogy network, that is a temporal directed network where each node represents a researcher and a directed link is a mentorship relation pointing from a mentor to their mentee (Fig. \ref{fig:schematic}a,b). Each node has a time attribute, corresponding to their doctoral or postdoctoral graduation year. The nodes included in this network are: (i) the node corresponding to the researcher $p$, (ii) the mentor of $p$, (iii) the set of nodes that are mentored by $p$'s mentor 5 years before and after $p$'s graduation, denoted as generation $G_1$, (iv) the set of nodes mentored by the nodes of $G_1$ during the first 25 years after graduation, denoted as generation $G_2$. For example, in Fig. \ref{fig:schematic}a, we show the academic genealogy network of researcher $p_1$.
To account for the longitudinal limits of the dataset, for each researcher we only consider two generations of nodes and include in $G_2$ only the mentees mentored during the first 25 years after a researcher's graduation (Supplementary Fig. S2 and Table S2). Also, we consider only researchers that graduated between 1900 and 1995, in order to have at least 25 years of career after graduation and to avoid right-censoring issues\cite{leung1997censoring}.

In order to understand the relation between the mentees' academic performance and the mentorship environment they were trained in, we introduce the concept of \textit{group size} and provide measures of \textit{academic performance}.
The group size of a given mentee is defined as the total number of nodes in $G_1$, that is the number of mentees that were supervised by the same mentor 5 years before and after the mentee's graduation. For example, in Fig. \ref{fig:schematic}a, the node $p_1$ is mentored with two other mentees during the five years before and after $p_1$'s graduation, whereas in Fig. \ref{fig:schematic}b, $p_4$ is mentored with four other mentees five years before and after $p_4$'s graduation. The group size is thus 3 for $p_1$ and 5 for $p_4$. Notice that the group size associated with a mentee is fixed, but a mentor can lead a group whose size can change over time and is equal to the number of mentees mentored in any 10 years window. The usage of this time years window ($d$ in Fig. \ref{fig:schematic}) to quantify the group size is motivated by previous work \cite{lienard2018intellectual}. We also show that group size has the same distribution when using different time windows (Supplementary Fig. S3), indicating that our findings do not depend on the choice of $d$.

Next, we define small groups and big groups: we first identify all the mentees who graduated in a given year, then we rank them in descending order according to their group size. Mentees in the top 25\% are in \textit{big} groups, while those in the bottom 25\% are in \textit{small} groups. Since the average and the 75\% quantile of group size are generally increasing over time \cite{wuchty2007increasing,wu2019large}, the threshold separating big groups and small groups is time-dependent, and accounts for the increasing size effect (Supplementary Fig. S3 and Table S3). 

\begin{figure}[!bt]
    \centering
    \includegraphics[width=1\textwidth]{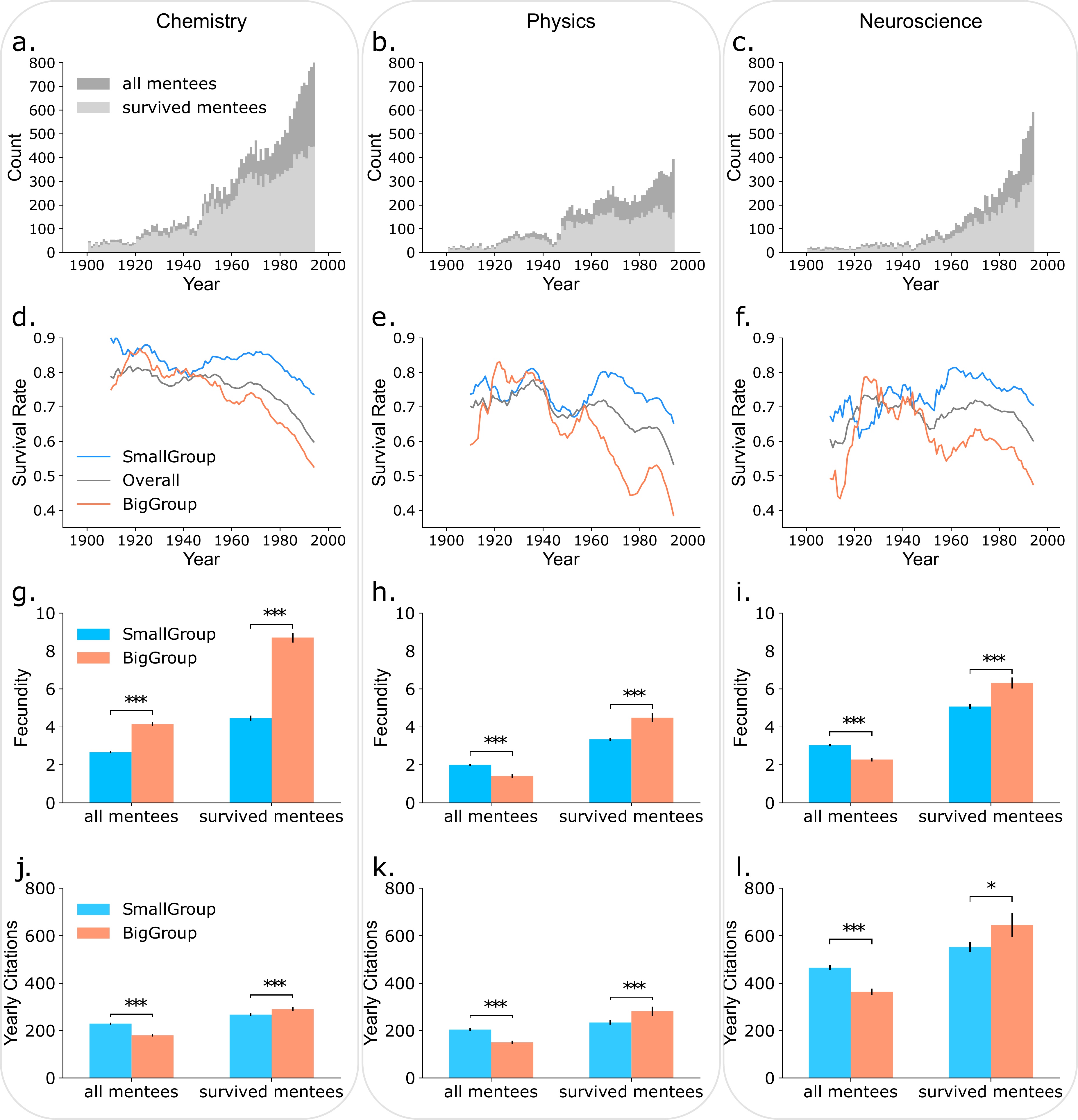}
    \caption{\label{fig:survival_rate}\textbf{Survival rate, fecundity, and yearly citations.} \textbf{a-c.} The evolution of the total number of mentees (dark grey bars) and the number of survived mentees (light grey bars). Survived mentees are those that had at least one mentee themselves. \textbf{d-f.} The evolution of the survival rate of mentees from small groups (blue lines) and from big groups (orange lines), compared with the overall average survival rate (grey lines). \textbf{g-i.} Fecundity during the first 25 years of career of all mentees (left two bars in each subplot) and of survived mentees (right two bars in each subplot) from small groups and big groups. \textbf{j-l.} Average yearly citations of all mentees (left two bars in each subplot) and of survived mentees (right two bars in each subplot) from small groups and big groups. The result of the Mann-Whitney significance test comparing distributions is reported at the top of each paired bars (*p < 0.05; **p < 0.01; ***p < 0.001).}
\end{figure}

\subsection*{Academic performance measures}
To quantify the academic performance of a mentee, we use three kinds of widely used measures \cite{malmgren2010role,milojevic2018changing,xing2019strong}: \textit{fecundity}, \textit{survival}, and \textit{publishing performance indicators}.
The \textit{fecundity} of a node is defined as the number of their mentees 25 years after graduation, that is the number of their neighbors in $G_2$. For example, the node $p_1$ trains two mentees in $G_2$ (Fig. \ref{fig:schematic}a), while $p_3$ has no trainees  (Fig. \ref{fig:schematic}b), hence $p_1$ and $p_3$ have fecundity equal to two and zero respectively.

Like previous work\cite{lienard2018intellectual}, we define \textit{survival} as having at least one mentee during the first 25 years, which is equivalent to having at least one neighbor in $G_2$. These researchers are ``surviving'' because they eventually establish themselves and build a scientific group. In contrast, mentees that do not have mentees themselves are considered dropouts, as they have not built a scientific group of their own, although they might continue publishing. In Fig. \ref{fig:schematic}a and b, the blue and green circles are survived mentees while grey circles are dropouts. We then define the \textit{survival rate} of the group, which is the fraction of nodes in $G_1$ that survive. For example, in Fig. \ref{fig:schematic}a 2 out of 3 mentees in $G_1$ survive, as they have a mentee in $G_2$, then the survival rate associated with this (small) group is 0.67. Similarly, in Fig. \ref{fig:schematic}b the survival rate of the (big) group is 0.4. We use also alternative definitions of survival\cite{milojevic2018changing,xing2019strong}, based on the mentee's publication record 10 years after graduation, and obtain findings similar to those shown in the following sections (Supplementary S2.2 and Fig. S6).

In addition to measures of academic survival and fecundity, we focus on publishing performance, as captured by \textit{productivity}, that is the number of papers published during the mentee's career, and \textit{average number of yearly citations} acquired by these papers. 
Finally, we use the publication record also to construct the \textit{collaboration} network between the mentor and all the mentees in $G_1$, to understand if this has an effect on the future career of mentees. In Fig. \ref{fig:schematic}c and \ref{fig:schematic}d, we provide two examples of collaboration networks, derived by the shown publications, where a node represents an author and two nodes are linked if they co-author at least one publication. The link weight corresponds to the number of co-authored publications.

\subsection*{Mentees trained in big groups have lower survival rate}
Given that fecundity and publications are both widely used as a proxy of success\cite{malmgren2010role,wuestman2020genealogical,rossi2017genealogical,clauset2017data}, we ask: what are the success differences between mentees from big groups and small groups? Who will perform better in their future academic career: Those trained together with many other mentees in supposedly high-profile large groups or those trained with just a few other mentees? Apart from group size, are there any other confounding factors associated with the development of a mentee's career?

To answer these questions, we first investigate the evolution of the total number of mentees (dark grey bars) and survived mentees (light grey bars) (Fig. \ref{fig:survival_rate}a-c). The total number of mentees has overall significantly increased from 1910 to 2000. In particular, after a temporary slow down soon after the World War II, the second half of the 20th century has witnessed a striking increase in both the total number of mentees and survived mentees, that continued until today. However, the rate of survived mentees was lower than the total number of mentees, indicated by the increasing difference between the dark grey and light grey bars. Indeed, the survival rate  (grey lines in Fig. \ref{fig:survival_rate}d-f), is (i) relatively stable for Chemistry until the 60s, (ii) suffered from a temporary decrease during World War II for Physics, followed by an increase probably because of the newly revived welfare after the war, which provided a large number academic positions in university and research institutes, and (iii) for Neuroscience had many ups and downs before and during World War II, and an increase until the early 70s. However, for all three disciplines the survival rate exhibits a striking declining trend after the 70s, which is still ongoing. When we split the survival rate into big groups and small groups, we find a pronounced difference (Fig. \ref{fig:survival_rate}d-f): Mentees from big groups have a significantly lower survival rate than those from small groups starting in the 1940's (Chemistry) or 1950's (Physics and Neuroscience), indicating that mentees from big groups were much less likely to continue in academia. In the 90s, the survival rate of mentees trained in big groups is between 30\% and 40\% lower than those from small groups. The different results about survival rates in small and big groups do not depend on the time-dependent threshold identifying small groups and big groups (Supplementary Table 3, Fig. S6-S7).  

\subsection*{Mentees trained in big groups have higher fecundity and higher yearly citations}
In the following analysis, we mainly focus on the data after 1960 when big groups and small groups exhibit evident difference in survival rate. We find significant differences between the mentees from big groups and small groups also in the other academic performance measures: Mentees from small groups are on average more successful in both fecundity (left two bars in the panels of Fig. \ref{fig:survival_rate}h-i) and yearly citations (left two bars in the panels of Fig. \ref{fig:survival_rate}j-l) than those from big groups, except for fecundity in Chemistry (Fig. \ref{fig:survival_rate}g). One possibility leading to this exception is that Chemistry is a predominantly experimental discipline requiring a large workforce, therefore the mentees from big groups inherit from their mentoring groups a much larger fecundity than those from small groups. Interestingly, when we compare the academic achievements of only survived mentees, that is mentees that have at least fecundity one, the observed performance differences in fecundity and average yearly citations reverse between groups (right two bars in each panel of Fig. \ref{fig:survival_rate}g-l). This means that mentees from small groups tend to do better in terms of average fecundity and yearly citations than those from big groups. However, if we consider only the mentees that manage to survive and establish a group, those from big group have an advantage, since they tend to have higher fecundity and yearly citations. Taken together, this advantage reversal happens because of the low survival rate in big groups, which lowers the average fecundity and citations of mentees from big groups. These findings are not a trivial consequence of dividing the data into the small groups and big groups, as shown by a null model where we randomize the mentor-mentee relationships while keeping the group size constant. In this null model, we do not see significant differences between mentees from big groups and small groups (Supplementary Fig. S4). The findings about survival suggest that being mentored in a big scientific group can have long-term career competitive advantages in academic performance, but these are conditional to the lower odds of surviving.

\begin{figure}[!b]
    \centering
    \includegraphics[width=1\textwidth]{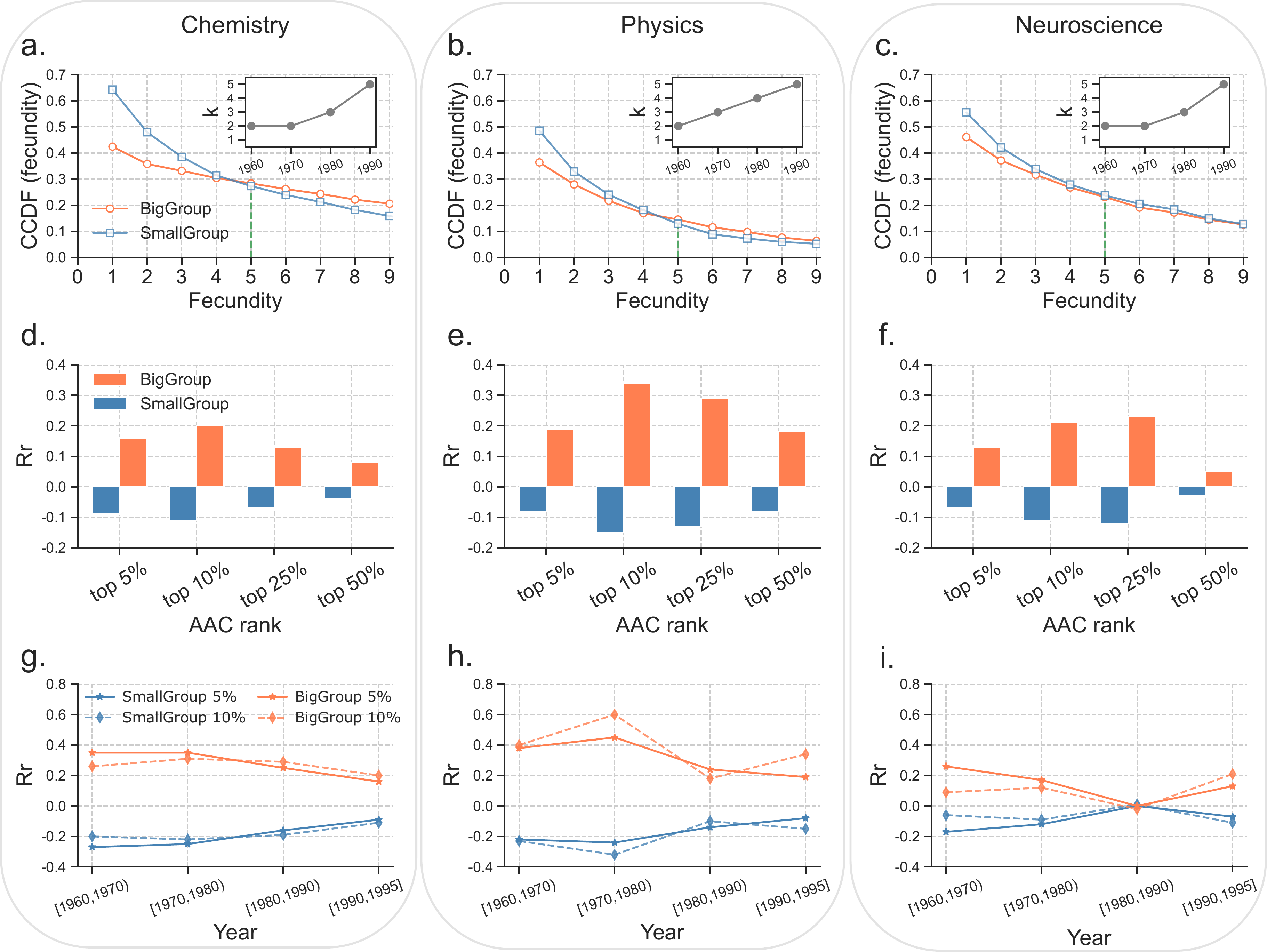}
    \caption{\textbf{Fecundity distribution and citation representation disparity among mentees from big groups and small groups.} \textbf{a-c.} Complementary cumulative distribution function (CCDF) of fecundity. The orange (blue) line displays the fecundity distribution of $G_1$ mentees from big (small) scientific groups. The green dashed line marks the point where the two probabilities are equal and the two distributions cross (only for Chemistry and Physics). \textbf{Inset:} The evolution of the equal probability point, $k$, for each decade since 1960s. \textbf{d-f.} Relative representation $Rr$ (see Methods) of $G_1$ mentees from small and big groups in the top 5\%, 10\%, 25\%, and 50\% of the average annual citations (AAC) ranking . \textbf{g-i.} The evolution of $Rr$ in the top 5\% and top 10\% of the AAC ranking.}
    \label{fig:evolution}
\end{figure}

\subsection*{Researchers trained in small groups have small groups, researchers trained in big groups have big groups}
The results in Fig. \ref{fig:survival_rate}g-l imply that big groups and small groups have different advantages based on the chosen success metric, namely survival probability, future fecundity, and average citations. Here we further explore the respective advantages of the two kinds of groups according to the academic aim of a mentee. We investigate the complementary cumulative distribution function (CCDF) of the mentees' fecundity, that is the probability that a researcher has at least $k$ mentees in their career, (Fig. \ref{fig:evolution}a-c), and focus on the value of $k$  where the probabilities for researchers trained in big groups and small groups are equal.
We observe that for Chemistry and Physics there is one point where the two probabilities are equal and two distributions cross. The point of crossover is at $k=5$ in the period 1990-1995, meaning that the likelihood to survive and have 5 mentees or less is higher for researchers trained in small groups. On the other hand, researchers trained in big groups have a higher likelihood to mentor 5 or more mentees in their careers, despite their lower odds of surviving. The two distributions do not cross over in Neuroscience and display only minor, although statistically significant differences, indicating that researchers trained in small groups have a slightly higher probability of having $k$ mentees, for all values of $k$ in the period 1990-1995. The point of equal probability $k$ identifies two different regimes: One regime where fecundity is smaller than $k$ and is associated with a higher likelihood to mentees trained in small groups; in the other regime, fecundity is larger than $k$ and is associated with higher likelihood to mentees trained in big groups. This opposite role of small and big groups regarding fecundity suggests two different strategies: a big group is to be preferred if a mentee aims at high fecundity, while a small group is to be preferred if the aim of a mentee is to avoid dropout, although the expected fecundity will be smaller. We calculate the points of same probability for each decade since 1960s (Fig. \ref{fig:evolution}a-c insets, and Supplementary Fig. S5) and find an increasing trend with time. This phenomenon indicates that researchers trained in a big group face high risks of dropout, if their aim is not a high fecundity.

\begin{figure}[!b]
    \centering
    \includegraphics[width=1\textwidth]{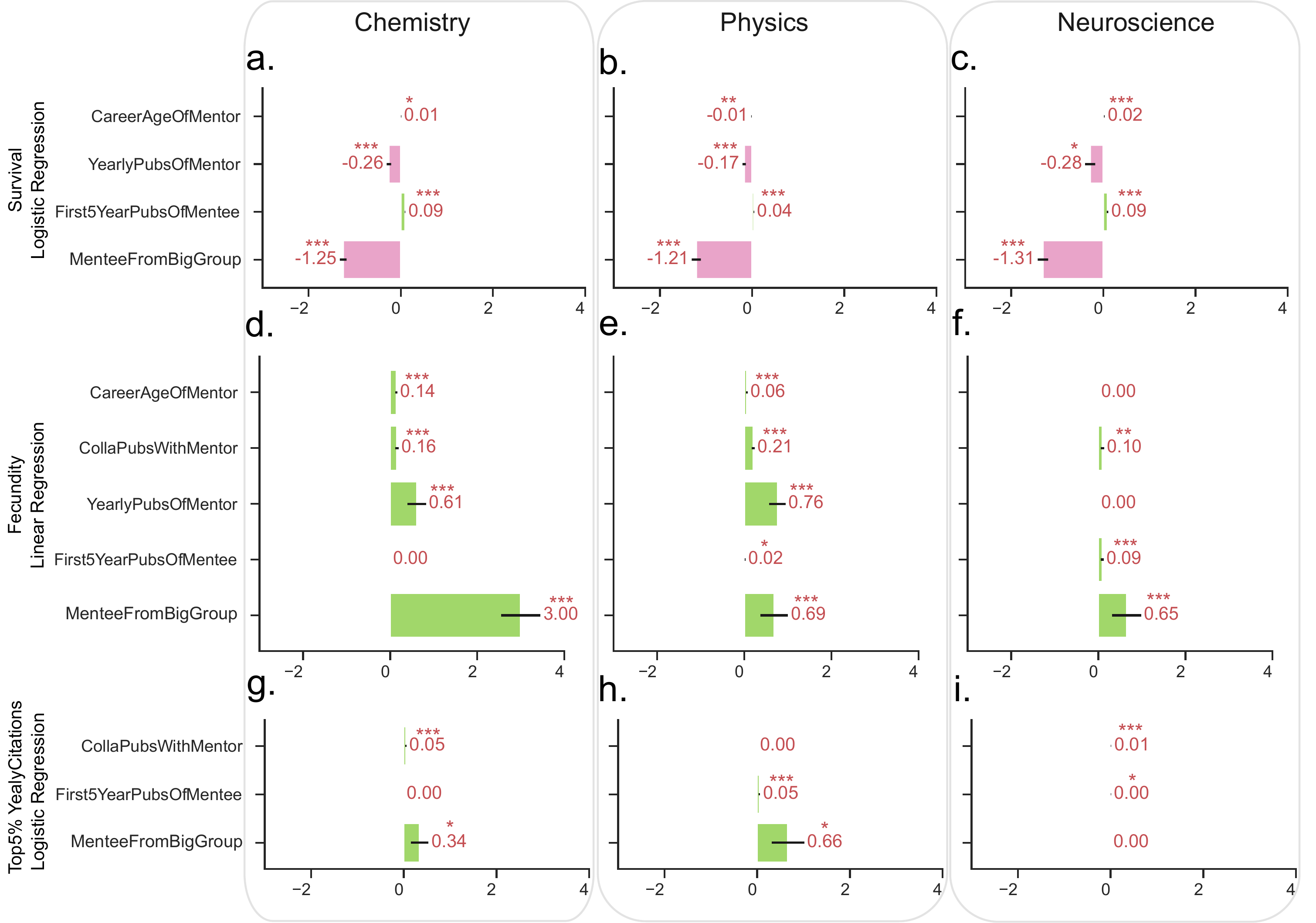}
    \caption{\textbf{Results of coarsened exact matching regressions.} \textbf{a-c.}
    Logistic regression of the odd of surviving in academia. The green (pink) bars indicate the positive (negative) regression coefficients for the corresponding variables on the y-axis. The numbers next to the bars indicate the value of the regression coefficient. The statistical significance of the variables are presented at the top of each value (* p < 0.05; **p < 0.01; ***p < 0.001). The error bars indicate the standard error of each regression coefficient. \textbf{d-f.} Linear regression of the fecundity of $G_1$ mentees. \textbf{g-i.} Logistic regression of the odds of being in the top 5\% in the AAC rank. Note that we show only the statistically significant variables of the regression models after coarsened exact matching the data.}
    \label{fig:regression}
\end{figure}
\subsection*{Big groups are more likely to nurture future top-cited researchers} 
Apart from academic fecundity, citation is one of the most popular and widely recognized metrics to measure a researcher's success. 
We measure the mentee's citation success by the probability of being a top-cited scientist. 
Based on previous work \cite{ma2020mentorship}, we measure the average annual citations (denoted by AAC) of each mentee during their career, and use it to create a ranking for each decade, based on the mentee's graduation year, from 1960 to 1995.  We then define the relative representation $Rr_{X\%}$ which captures how many mentees we observe in the top $X\%$ of the ranking compared to a random model where group size has no effect on the ranking (see Methods). In general, $Rr>0$ means that the mentees of a given group are more represented than expected. Conversely, $Rr<0$ indicates that mentees are underrepresented compared to the expectation. In Fig. \ref{fig:evolution}d, we consider the mentees trained in big groups and small groups in the period 1990-1995 and study their relative representation in the top 5\%, 10\%, 25\% and 50\% of the AAC ranking. We find that mentees from big groups, if surviving, are over-represented among top-cited scientists. Moreover, the result is more pronounced in the top 10\% and the pattern is consistent across different research fields. Taken together, Fig. \ref{fig:survival_rate}g-l and Fig. \ref{fig:evolution}d-f show that survivors from big scientific groups are not only likely to have a better average academic performance, but also have a competitive advantage in being top-cited scientists. We additionally study how the relative representation evolved over time (Fig. \ref{fig:evolution}g-i). Big groups are becoming less dominant in raising top-cited scientists in Chemistry and Physics in recent decades, as indicated by the orange line decreasing from 0.35 to 0.16 in Fig. \ref{fig:evolution}g and from 0.38 to 0.18 in Fig. \ref{fig:evolution}h. The same trend was present from 1960 to 1990 in Neuroscience, but seems to have changed in the 1990s (Fig. \ref{fig:evolution}i).
Survived mentees from small groups have become more represented than previously, even though they are still underrepresented compared to those surviving from big groups. However, we do not find evident changing trends with respect to the top 25\% and top 50\% AAC rank (Supplementary Fig. S8, S9 and S10 for details). Our results imply that the candidates surviving in a big group perform better in terms of impact than those from small groups. 

\subsection*{Controlling for confounding factors}
\begin{figure}[!bt]
    \centering
    \includegraphics[width=1\textwidth]{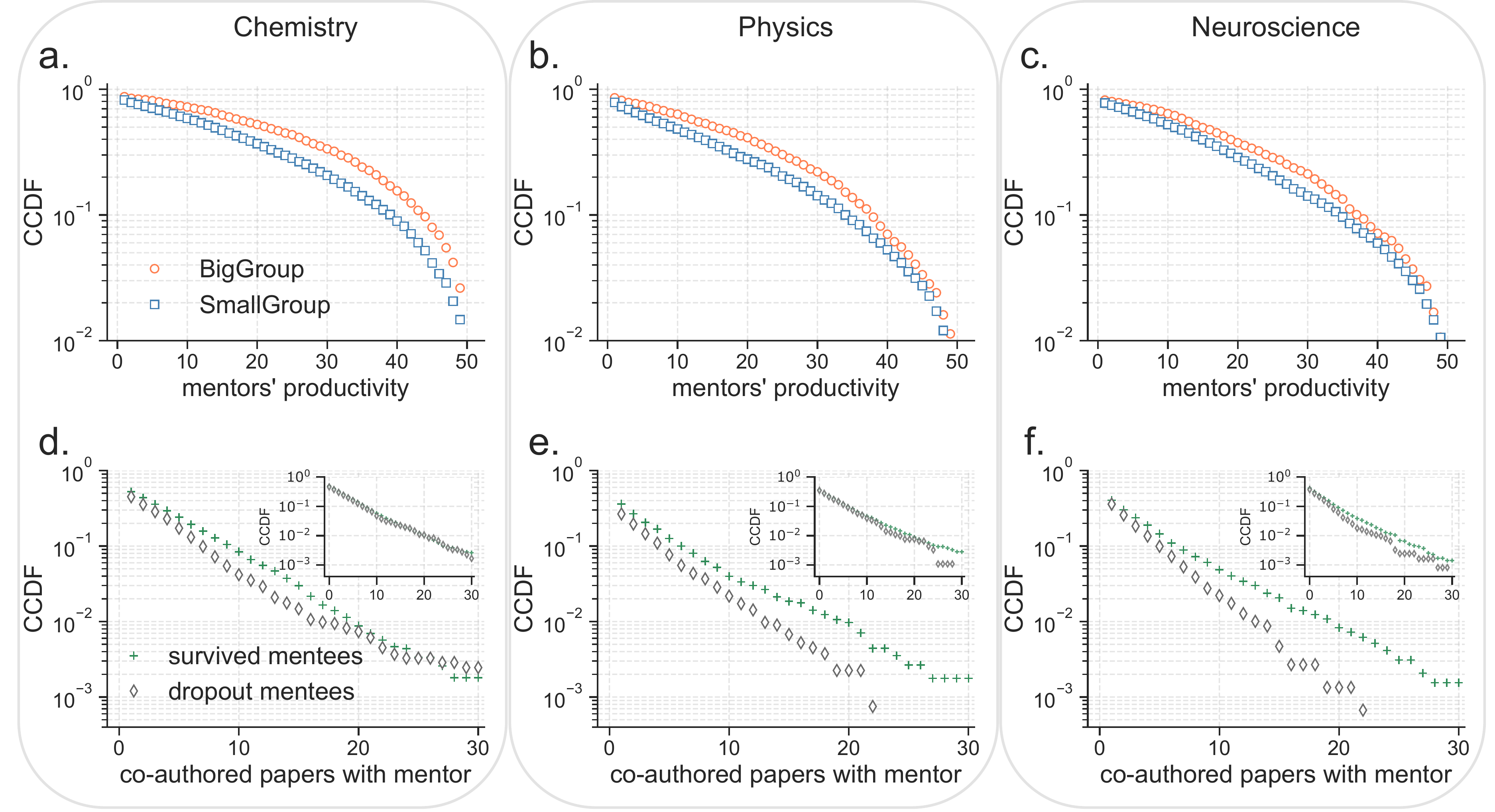}
    \caption{Mentors' productivity and collaboration with their mentees during the mentees' training period. \textbf{a-c.} The complementary cumulative distribution function (CCDF) of the mentors' productivity leading small groups (blue) and big groups (orange). Here productivity is the number of publications published by the mentor during the training period $d$, highlighted in Fig. \ref{fig:schematic}a. \textbf{d-f.} CCDF of the number of papers co-authored with the mentor by survived mentees (green pluses) and by dropped out mentees (grey diamonds). This data refers only to mentees trained in big groups. \textbf{Inset:} CCDF of the number of papers co-authored with the mentor of survived mentees (green pluses) and dropped out mentees (grey diamonds) trained in small groups only.}
    \label{fig:regression_support}
\end{figure}

In order to understand the role of potential confounding factors, we use a coarsened exact matching (CEM) coupled with regression models to study the relation between scientific group size and predictors of future academic performance. CEM regression consists in running a separate regression model on matched groups of mentees, resulting in a more stringent way of controlling for confounding factors than regression alone (Methods)\cite{iacus2009cem,iacus2012causal}. In Fig. \ref{fig:regression}a-c, the logistic regression applied to CEM datasets 
shows that the most significant variable, with a negative weight, to predict survival is \textit{MenteeFromBigGroup} variable, confirming the finding that being trained in a big group lowers the odds of future survival in academia. The positive regression coefficient of the variable \textit{First5YearPubsOfMentee} indicates that a mentee’s early productivity is associated with survival, supporting previous findings \cite{milojevic2018changing,xing2019strong}. Moreover, working with senior supervisors (larger \textit{CareerAgeOfMentor}) rather than with junior supervisors gives a slight yet significant advantage to survive in Chemistry and Neuroscience, while it seems to have a slight negative effect on mentee survival in Physics. Also, the regression coefficient of the \textit{YearlyPubsOfMentor} variable indicates that the mentor's yearly productivity, i.e. the average number of papers published in a year, has a negative effect on the mentee's survival probability. Taken together, a possible explanation for the results observed in Fig. \ref{fig:regression}a-c is that busy mentors, such as those from big groups and with a high publishing rate, have typically little time to spend on supervising each mentee, affecting their future academic career. 
The negative association between mentor productivity during the mentee training and mentee success is further confirmed when we study the distribution of the number of mentor's papers divided by group size (Fig. \ref{fig:regression_support}a-c). The CCDF for mentors supervising big groups is always larger than those supervising small groups, indicating that mentors leading big groups tend to have more publications on average than those leading small groups. At the same time, their mentees have a lower survival probability than mentees trained in small groups (Fig. \ref{fig:survival_rate}d-f).

We use a regression approach on CEM datasets also to control for confounding factors in the prediction of fecundity and citation performance, and confirm our previous observations: Group size is a significant factor, being positively associated with future fecundity and citation performance, captured by being among the top 5\% scientists for yearly citations (\textit{Top5\%YearlyCitations}, Fig. \ref{fig:regression}d-i). The only exception is Neuroscience, where group size is not significant to predict a top-cited scientist (Fig. \ref{fig:regression}j).
Overall, the regression analysis confirms that, if surviving, a mentee from a big group  has long-term competitive advantages compared to small groups. 

Apart from group size, we find one more variable, the number of papers co-authored with a mentor during training (\textit{CollaPubsWithMentor}, see schematic illustration in Fig. \ref{fig:schematic}c-d), which is positively associated with fecundity and citation performance. This is compatible with the hypothesis that the mentees that receive more supervision from their mentors, as signalled by the higher number of coauthored papers, have higher chances of future success.
This is further confirmed by the observed statistical disparities of the mentor-mentee collaboration in big groups and small groups (Fig. \ref{fig:regression_support}e-g): in big groups, mentees who will survive tend to have more co-authored papers with a mentor during training than those that will dropout. For mentees trained in small groups, there is no noticeable distribution difference between survived mentees and dropepd out mentees  (Fig. \ref{fig:regression_support}e-g Insets). Mentees working in small groups can receive more even attention from the mentor in the same period because there are fewer trainees. Finally, our regression models have between $66\%$ and $73\%$ prediction accuracy in mentee survival and reveal another main factor (Fig. \ref{fig:regression} and Supplementary Table S5, Fig. S15): Surprisingly, the more productive a mentor is, the smaller the probability that their mentee will stay in academia.Taken together, our findings quantitatively support the hypothesis that the attention received from the mentor plays a key role for the higher survival rate and success of mentees in academia\cite{ma2020mentorship}.

\section*{Discussion, limitations, and conclusions}\label{sec:discussion}
Our findings about the effects of group size and mentor productivity support our hypothesis that the attention allocation of the mentor affects the future academic success of mentees: A highly productive mentor supervising a big group tends to provide less supervision opportunities to each mentee, which results in a higher dropout rate. In big groups, this tendency is counterbalanced only when there are frequent collaborations, and hence more supervision opportunities, between mentee and mentor. Taken together, based on large-scale data in scientific genealogy and scientometrics, we offer empirical evidence for both potential profits and risks of working with successful mentors. 

Our study has some limitations. First, some mentorship relations might not be reported in the AFT dataset, which could affect the actual group size measure. To mitigate this issue, we analyzed only the three most represented fields in the data: Chemistry, Neuroscience, and Physics (Supplementary S1.2). Our CEM and regression analysis should also mitigate reporting bias due to the different visibility of mentors, since we control for individual productivity and citations. Also, prior literature has widely investigated the AFT dataset\cite{lienard2018intellectual,ma2020mentorship,schwartz2021impact,david2012neurotree,ke2022dataset}, and has not found obvious biases that could affect our findings. Second, the AFT dataset only reports formal mentorship relations. In academia, graduate students receive informal mentorship from many researchers, including postdocs, teachers, other faculty and academic staff \cite{acuna2020some}. These relations are not captured by this data set and, to our knowledge, by no other openly available data set. Yet, while information about informal mentorship could provide more causal explanation to our findings on career evolution, the reported relation between group size of the official mentor and survival, fecundity, and academic achievements would still hold. Third, in this study we use a narrow definition of academic achievements, such as survival, fecundity, and average annual citations. These measures are oblivious of other dimensions of success, not quantifiable in our data, and do not fully represent a successful academic career, in all its aspects. Yet, since decisions in the academic enterprise are lately strongly driven by quantitative measures like those used in this paper, we believe that it is important to study the properties and relations between these indicators.    

Our findings indicate that a simple characteristic such as the size of the mentor's group can help predicting the long-lasting achievements of a researcher's career. Our work also raises important questions: Should research policies balance the number of mentees per mentor, given the association with a higher dropout rate? Or should they promote excellence of future career, as arguably nurtured in big successful groups, despite the higher risk of dropout?  
There are also open questions that we have not tackled here but that offer important future directions of inquiry. 
An important one is: what is the effect of group structure on the mentees' success? We have shown a strong collaboration between mentee and mentor counterbalances the lower odds of survival in a big group. However, we have not explored the role of the inner group structure, as captured by collaboration ties between the supervised mentees or between a mentee and other junior academics. These collaborations could provide mutual support, mitigating future dropout risk.
Other important open questions concern how our findings change when differentiating data based on gender, country of origin, or ethnicity. Indeed, previous research shows the existence of strong biases in mentorship and in science in general \cite{moss2012science,lariviere2013bibliometrics,dutt2016gender,dennehy2017female,schwartz2021impact,hernandez2020inspiration} which could intersect in problematic ways with the big group effect. Answering these questions could not only offer a better understanding of the fundamental mechanisms that underpin a scientific career from the beginning but might also substantially improve our ability to retain young researchers, to improve workplace quality, and to nurture high-impact scientists.

\clearpage

\section*{Methods}\label{sec:methods}
\subsection*{Data Preparation}
The Academic Family Tree (AFT, \url{Academictree.org}) records formal mentorship mainly based on training relationships of graduate student, postdoc and research assistant from 1900 to 2021. AFT includes 743,176 mentoring relationships among 738,989 scientists across 112 fields. The data can be linked to the Microsoft Academic Graph (MAG, \url{https://aka.ms/msracad}) which is one of the largest multidisciplinary bibliographic databases. The combined data contains the publication records of mentors and mentees, which can be used to calculate the measurements of publication-related performance in our analysis (Supplementary Note 2 and Note 3). The combined data of AFT and MAG is taken from \cite{ke2022dataset}. In this paper, we conduct our analysis on researchers in Chemistry, Physics, and Neuroscience, amounting to 350,733 mentor-mentee pairs, and to 309,654 scientists who published 9,248,726 papers. We motivate our choice for the studied fields in Supplementary Note 1 (Data and preprocessing). 

\subsection*{Relative Representation (Rr)}
Given a time window, we rank the mentees graduated in this window according to their average annual citations (AAC), calculated over their whole career. Then we compute the observed representation of big group mentees, $R_{BG}(X)$ in the top X\% of the AAC ranking as:
\begin{equation}
\label{eq1}
        R_{BG} (X) = \frac{N_{BG}(X)}{N_{BG}(X) + N_{SG}(X)}    
\end{equation}
where $N_{BG}(X)$ and $N_{SG}(X)$ are the number of mentees from big groups and small groups, respectively, found in the top $X\%$ AAC ranking.
We compare this observed representation with the expected representation if the position in the ranking is independent of the group size a mentee is from. The expected representation is:
\begin{equation}
\label{eq2}
    R^{exp}_{BG}=\frac{N_{BG}}{N_{BG}+N_{SG}}
\end{equation}
where $N_{BG}$ and $N_{SG}$ are the total number of mentees from big groups and small groups, respectively.
The relative representation in the top $X\%$ ranking, $Rr_{BG}(X\%)$ is obtained by subtracting (\ref{eq2}) from (\ref{eq1}) and dividing by (\ref{eq2}):
\begin{equation}
\label{eq3}
   Rr_{BG}(X\%) = \frac{R_{BG} (X) - R^{exp}_{BG}}{R^{exp}_{BG}}    
\end{equation}
Similarly, the relative representation for small groups is defined as:
\begin{equation}
\label{eq4}
   Rr_{SG}(X\%) = \frac{R_{SG}(X) - R^{exp}_{SG}}{R^{exp}_{SG}}    
\end{equation}
where $R_{SG}(X)$ and $R^{exp}_{SG}$ are obtained by swapping $N_{BG}(X)$ and $N_{SG}(X)$ in (\ref{eq1}) and (\ref{eq2}). 

\subsection*{Coarsened Exact Matching (CEM) Regression}
In causal inferences, analyzing matched data set is generally less model-dependent (i.e., less prone to the modeling assumptions) than analyzing the original full data\cite{iacus2012causal,ho2007matching}. For this reason, we use a matching approach before applying regression models to our datasets. With a matching approach\cite{iacus2009cem,iacus2012causal}, two groups can be balanced, resulting in similar empirical distributions of the covariates. There are many approaches to matching: one approach is based on exact matching, which is the most accurate but also not usable in practice as it returns too few observations in the matched samples. Here, we use the Coarsened Exact Matching (CEM): this method first coarsen the data into linear bins, then matches elements of two groups that fall within the same bin. This approach returns approximately balanced data and allows to control for covariates. 
Taken together, the CEM approach involves three steps:
\begin{itemize}
  \item [1.] Given each mentee $i$, we define a vector $\mathbf{V}_i$ where each element of the vector corresponds to an individual variable, like number of publications or number of collaborators. 
  \item [2.] We coarsen each control variable, creating bins for each quantile of the distribution\cite{iacus2012causal} (Supplementary S3.2). Then for each $i$, we convert the vector $\mathbf{V}_i$ into a coarsened vector $\mathbf{V^C}_i$, where each element maps the individual variable to the corresponding bin of the coarsened variable.
  \item [3.] We then perform an exact matching of the coarsened vectors, that is for each $i$ in one group we find a $j$ in the other group such that $\mathbf{V^C}_i==\mathbf{V^C}_j$.
\item [4.] We discard all elements $i$ that we are not able to match.
\end{itemize}
These procedure returns to CEM datasets. After creating these datasets , we apply regression models to estimate the effect of the independent variable (group size) on outcome variables (survival, fecundity, and yearly citations). 
Specifically, we use a logistic regression, a linear regression, and a logistic regression model to study the effects of group size, respectively, on the mentee's survival, fecundity, and being among the top5\% cited researchers. See the Supplementary Note 3 and Note 4 for details on variables, CEM regression models, Chi-square test and cross-validation.

\subsection*{Regression variables}
The regression includes controls of the following variables: 
\begin{itemize}
  \item [$\bullet$]\textit{YearlyPubsOfMentor} -- number of yearly publications over a mentor’s career.
  \item [$\bullet$]\textit{TotalPubsOfMentor} -- number of total publications over a mentor’s career.
  \item [$\bullet$]\textit{YearlyCitationOfMentor} -- number of yearly citations over a mentor’s career.
  \item [$\bullet$]\textit{TotalCitationOfMentor} -- number of total citations over a mentor’s career.
  \item [$\bullet$]\textit{YearlyCollaOfMentor} -- number of yearly coauthors over a mentor’s career.
  \item [$\bullet$]\textit{TotalCollaOfMentor} -- number of total coauthors over a mentor’s career.
  \item [$\bullet$]\textit{PubsOfMentorInTraining} -- number of mentor’s papers during a given mentee's training period.
  \item [$\bullet$]\textit{CareerAgeOfMentorInTraining} -- a mentor’s career stage at the mentee's graduation.
  \item [$\bullet$]\textit{First5YearPubsOfMentee} -- number of publications in the first 5 years of a mentee's career.
  \item [$\bullet$]\textit{First5YearCitationOfMentee} -- number of total papers' citations in the first 5 years of a mentee's career.
  \item [$\bullet$]\textit{First5YearCollaOfMentee} -- number of coauthors in the first 5 years of a mentee's career.
  \item [$\bullet$]\textit{CollaPubsWithMentor} -- number of co-authored papers with their mentor during training period.
  \item [$\bullet$]\textit{MenteeFromBigGroup} -- the independent variable is 1 or 0 if the mentee graduated from a big group or small group in the regressions.
  \item [$\bullet$]\textit{survival} -- the dependent variable is binary.
  \item [$\bullet$]\textit{fecundity} -- the dependent variable is discrete.
  \item [$\bullet$]\textit{Top5\%YearlyCitations} -- the dependent variable that is being among the top 5\% scientists for yearly citations, which is also binary.
\end{itemize}
More information about the variables of the regressions can be found in the Supplementary Note 3.

\bibliography{groupsize}

\section*{Acknowledgements}

We thank all members of the Networks, Data, and Society (NERDS) research group at IT University of Copenhagen, and especially Michael Szell, for their helpful discussions. This work is supported by the National Natural Science Foundation of China under Grant (Nos. 71731002). Y.X. acknowledges the support from China Scholarship Council. R.S. acknowledges supports by Villum Fonden through the Villum Young Investigator program (project number: 00037394). 

\section*{Author contributions statement}
Y.X. and A.Z. conceived the study. All authors contributed to the design of the study. Y.X. curated the datasets. Y.X., R.S. and A.Z. performed the analysis. Y.X., R.S. and A.Z. contributed to the interpretation of the results and writing of the manuscript. R.S. was the lead writer of the manuscript.

\section*{Additional information}

\textbf{Accession codes and data} The codes and curated dataset can be acquired from the first author Y.X..\\
\textbf{Supplementary Information}\\
\url{https://drive.google.com/file/d/1lhBoi3zKTU0w8z1MafGofPdqvXz9DGWe/view?usp=sharing}\\
\textbf{Competing interests} The authors declare no competing interests.

\end{document}